\documentclass[manuscript]{aastex}

\slugcomment{Submitted to the Astrophysical Journal, October 18, 2001.
  Resubmitted to
the Astrophysical Journal, April 18, 2002.}
\shorttitle{Intracluster Magnetic Fields}
\shortauthors{Newman et al.}

\begin{document}

\title{Quantification of Uncertainty in the Measurement of \\
Magnetic Fields in Clusters of Galaxies}

\author{William I. Newman}
\affil{Departments of Earth \& Space Sciences, Physics \& Astronomy, and
Mathematics, University of California, Los Angeles, CA 90095 ,\\and\\
Department of Computer Science \& Applied Mathematics, Weizmann
Institute of Science, Rehovot 76100, Israel}
\email{win@ucla.edu}

\author{Alice L. Newman}
\affil{Department of Physics, California State University at Dominguez
Hills, Carson, CA 90747, \\and\\
School of Physics \& Astronomy, Tel Aviv
University, Tel Aviv 69978, Israel}
\email{anewman@csudh.edu}

\and

\author{Yoel Rephaeli}
\affil{School of Physics \& Astronomy, Tel Aviv University, Tel
Aviv 69978, Israel
, \\ and\\
Center for Astrophysics and Space Sciences,
University of California, San Diego,\\ La Jolla, CA 92093}
\email{yoelr@wise1.tau.ac.il}

\begin{abstract}
We assess the principal statistical and physical uncertainties associated
with the determination of magnetic field strengths in clusters of galaxies
from measurements of Faraday rotation (FR) and Compton-synchrotron
emissions. In the former case a basic limitation is noted, that the
relative uncertainty in the estimation of the mean-squared FR will
generally be at least one third. Even greater uncertainty stems from the
crucial dependence of the Faraday-deduced field on the coherence
length scale characterizing its random orientation; we further elaborate
this dependence, and argue that previous estimates of the field are
likely to be too high by a factor of a few. Lack of detailed spatial
information on the radio emission---and the recently deduced nonthermal
X-ray emission in four clusters---has led to an underestimation of the
mean value of the field in cluster cores. We conclude therefore that it
is premature to draw definite quantitative conclusions from the
previously-claimed seemingly-discrepant values of the field determined by
these two methods.
\end{abstract}


\section{Introduction}

Spectral and imaging observations of clusters of galaxies with the
{\it  XMM} and {\it Chandra} satellites at energies $\epsilon \leq 10$
keV, are currently yielding detailed information on the thermal
properties of the hot intracluster (IC) gas. The improved determinations
of gas  temperature, density, and metal abundances  will greatly improve
our knowledge of the intrinsic properties ({{\it e.g.},\,} total mass) of clusters,
and will significantly  advance the use of clusters as more valued
cosmological probes ({{\it e.g.},\,} for the measurement of the Hubble constant,
$H_0$). However, as has been the case in galaxies, in
clusters too a more complete description of the astrophysics of these
systems necessitates knowledge of basic nonthermal (NT) quantities such
as magnetic fields, and the properties of relativistic electrons and
protons.

Direct evidence for the presence of relativistic electrons and magnetic
fields in clusters comes from measurements of extended regions of radio
(in the  frequency range $\sim 0.04-4$ GHz) synchrotron emission in
$\sim 30$ clusters (see Giovannini et al.\, 1999, Giovannini et al.\, 2000,
and references therein). In many of the clusters the emitting
region is central, with a typical size of $\sim 1-3$ Mpc.
Values of the mean magnetic field and relativistic electron energy
densities can be obtained if an assumption is made concerning the
relative energy densities in the particles and field. Mean field values
of a few $\mu$G have been obtained under the assumption of global energy
equipartition. Magnetic fields have also been deduced from the measurement
of FR of the plane of polarization of distant radio sources seen through
clusters ({{\it e.g.},\,} Kim et al.\, 1991). The observed drop in the
absolute value of the RM over a typical cluster-centric distance of
$\sim 0.5$ Mpc provides a statistical measure of the mean field along
the line of sight to the radio source. Specifically, in a recent survey
of radio sources seen through the cores of 16 clusters, Clarke,
Kronberg, and B\"ohringer (2001, hereafter CKB) have deduced a mean
field value of $\sim 5-10\, (\ell/10\ {\rm kpc})^{-1/2}$ $\mu$G, where
$\ell$ is a characteristic field spatial coherence (reversal) length.

An important recent development is the apparent detection of NT X-ray
emission in four clusters, namely Coma, A2199, A2256, and A2319, by the
RXTE (Coma cluster: Rephaeli, Gruber, \& Blanco 1999, hereafter RGB;
A2319: Gruber \& Rephaeli 2001) and by the BeppoSAX (Coma: Fusco-Femiano
et al.\, 1999; A2256: Fusco-Femiano et al.\, 2000; A2199: Kaastra et al.\, 2000)
satellites. The most likely origin of NT X-ray emission in clusters is
Compton scattering of relativistic electrons by the cosmic microwave
background (CMB) radiation (Rephaeli 1979). The radio and NT X-ray
emissions in three of these clusters are likely to be related, providing
considerable motivation for their combined analysis. If so, then the
mean value of the magnetic field and relativistic electron density can
be extracted based directly on observables---the radio synchrotron and
Compton-produced fluxes and the joint power law index. Of the four
clusters, the best studied is Coma, for which both RGB and Fusco-Femiano
et al.\, (1999) determine a mean field value of $\sim 0.2$ $\mu$G.

Emission by NT electrons may possibly be detected also at lower
energies: EUV observations of several clusters have reportedly led to
the measurement of diffuse low-energy ($65-245$ eV) emission which is
possibly NT (Sarazin \& Lieu 1998, Bowyer \& Berghofer 1998). However,
Bowyer et al.\, (1999) argue that this emission has been unequivocally
detected {\it only} in the Coma cluster. The origin of the EUV emission
is uncertain; suggestions range from NT bremsstrahlung by a population
of energetic (or `suprathermal') electrons which is separate from the
radio-emitting relativistic \, electrons) ({{\it e.g.},\,}, Sarazin \& Kempner 2000), to
Compton scattering of the CMB by lower energy relativistic \, electrons (Ensslin,
Lieu \& Biermann 1999). Because of the very substantial uncertainties
regarding the nature and origin of the observed EUV emission, we consider
here only the relativistic \, electron population that is directly deduced from
radio and possibly also NT X-ray emission.

Magnetic field values from Compton-synchrotron, $B_{rx}$, or FR,
$B_{fr}$, observations clearly involve very different spatial averages
not only of the field itself, but also of the relativistic electron
density, or the gas density, along the line of sight, respectively.
Whereas $B_{rx}$ is based on a volume average of the magnitude
of the field over typical radial regions of $\sim 1$  Mpc, $B_{fr}$
is a weighted average of the field vector {\it and} gas density
along the line of sight. Generally, therefore, it is not expected that
these two measures of the field yield similar values. Indeed, it has
already been shown that $B_{rx}$ is typically expected to be smaller
than $B_{fr}$ (Goldshmidt \& Rephaeli 1993). Moreover, the full
expression for $B_{rx}$ includes a ratio of spatial factors (Rephaeli
1979) which are essentially volume integrations over the spatial
profiles of the relativistic electrons and field. For lack of spatial
information on these profiles, this ratio has been taken to be unity
in the analyses of RXTE and BeppoSAX data, resulting in a systematically
lower value of $B_{rx}$.
Generally, FR observations yielded only a measure of the average field
over a {\it sample} of clusters; this makes a comparison between $B_{fr}$
and $B_{rx}$ somewhat uncertain. Such a comparison is more meaningful in
the case of the Coma cluster, for which there are specific measurements
of both $B_{fr}$ (Kim et al.\, 1990, Feretti et al.\, 1995) and $B_{rx}$ (RGB,
Fusco-Femiano et al.\, 1999), i.e., A400 and A2634 (Eilek \& Owen 2001).

Evidently, the interpretation of cluster Compton-synchrotron and FR
observations is not straightforward, particularly with regard to the
measurement of the strength of IC fields. In addition to observational
errors, there are substantial uncertainties stemming from physical as
well as statistical considerations whose effects have never been fully
investigated. In view of current and near-future observational
capabilities, a systematic study of all the relevant issues is both
desirable and timely, and the main objective of this paper.

Large uncertainties in the interpretation of FR, radio synchrotron, and
Compton emissions stem from the small number of background radio sources
(for FR measurements), and poor (radio) or no (X-ray) spatial information on the
distribution of NT emission. The main cause of error is the likely
complex morphology of IC magnetic fields and the energetic electron
density distribution. The fields are quite likely turbulent, polarized
over a range of coherence scales ($\sim 1- 50$ kpc), and with large scale
($>100$ kpc) variation of their mean strength. Additional sources of
uncertainty are due to the small number of bright radio sources behind
galaxy clusters that can serve as probes along their respective lines of
sight for estimating polarization effects due to the magnetic field.
While increased detector sensitivity and imaging capability at high
($>20$ keV) X-ray energies will help mitigate these limitations, we
shall show that prevailing methodologies---despite great efforts to
minimize instrument error---remain incapable of reducing the uncertainty
in the magnetic fields, and that these uncertainties could be very
substantial indeed.  The influence of embedded, extended sources requires the development
of a more complex model which introduces their influence upon the surrounding
IC medium and field environment; this will be the subject of a future paper.

In the following section, we focus on statistical aspects emerging
from the determination of the FR measure in the ideal situation wherein
the IC medium and its magnetic field are both homogeneous in character.
We first show that there is a basic statistical limitation in our
ability to measure the mean-squared excess FR in clusters. Next, we
explore---in the spirit of Goldshmidt \& Rephaeli (1993)---the role of
inhomogeneity and turbulent field structure and how they are manifested
in observations of FR. Then, we briefly examine how the
differing spatial distributions of relativistic particles and fields
can affect our inference of field values. In the Discussion we assess
the overall quantitative measure of uncertainty in estimating field
strengths in clusters, and briefly consider some of its consequences.

\section{Statistical Issues in the Measurement of Faraday Rotation}

Theoretical research (Crusius-W\"atzel et al.\, 1990, Goldshmidt \& Rephaeli
1993) dealing with the FR measure has focused on the random nature of
the rotation angle $\chi$ defined by
\begin{equation}
\chi = c_1 \lambda^2 \int_0^s {\rm d}s\,
n_t \left( s \right) \,B_\parallel \left( s
\right)
\end{equation}
where $c_1=0.81$ if $\chi$ is measured in radians and $n_t$---the
thermal electron density---in $\rm cm^{-3}$; $B_\parallel$ is the
random magnetic field component in $\mu$G along the line-of-sight, and $s$
is  the path length in parsecs. Since $n_t\,B_\parallel$ behaves as a
random variable, the Central Limit Theorem (Feller 1968) and its
generalization, known as the Feller-Lindberg Theorem, apply and assure
that the outcome is well-approximated by a Gaussian with zero-mean
\footnote{The Central Limit Theorem is applicable
only in situations where the random variable which is being
``summed''---or integrated in our case---has statistically homogeneous
properties. If,  on the other hand, the underlying random variable has
systematic  properties, such as a power-law character in its variance (as
might be  expected from a spatial dependence of a magnetic field that
follows from  a King profile), then that integral will also appear to be
normally  distributed owing to the Feller-Lindberg Theorem.}.  We have
verified that the data presented by CKB, namely column 5 of  their Table
1, are Gaussian distributed.

    Crusius-W\"atzel et al.\, \ (1990) and Goldshmidt \& Rephaeli (1993)
suggested that the mean-squared fluctuation of the rotation angle be
employed instead, namely
\begin{equation}
\left\langle \chi^2 \right\rangle = c_1^2
\lambda^4 \int_0^s \int_0^s {\rm d}s' {\rm
d}s'' n_t \left( s' \right) n_t \left( s''
\right) \left\langle
B_\parallel \left( s' \right)
B_\parallel \left(s'' \right) \right\rangle
\end{equation}
where the mean of the rotation angle is presumed to vanish, i.e.,
$\left\langle \chi \right\rangle =0$, and where
$\left\langle ... \right\rangle$ denotes the expectation value operator.
In observational work by CKB that explored the statistical
nature of estimates of $\chi$, they presented data from 16 clusters of
galaxies that were assumed to have similar morphologies,
and analyzed the emission from 27 radio sources from which FR could be 
derived. Of the 27 sources projected in the 16 clusters in their sample, 
15 are embedded (Clarke, personal communication) cluster members, a fact 
that complicates the interpretation of their results. (We explore possible 
ramifications of this in the Discussion.) 
However, we will use the CKB sample as though it contained {\it only}\/
background point sources, such as quasars, as an observational basis for 
comparison with our Monte Carlo investigations.
A basic result from 
the work of CKB 
is the confirmation of the role of spatial inhomogeneity. In particular, 
they observed a clear excess FR in radio sources
viewed through the magnetized IC gas as compared to those viewed beyond 
the main gaseous region of the cluster. However, no specific information 
on the role of spatial inhomogeneity resulted from their statistical study, 
an issue addressed theoretically by Goldshmidt \& Rephaeli (1993). Indeed, 
an outstanding issue is how the position of the background source relative
to the center of the cluster, what CKB call the ``impact
parameter'', influences the excess FR.

    Since many researchers have invested substantial effort in the
determination of the excess FR, we address the statistical significance
of the estimates that can be derived from such studies. Suppose that we
have $N$ estimates of the Faraday rotation angle 
$\chi_i, \, i= 1, ...,  N$, such that the related impact parameters are 
sufficiently close to each  other, so that spatial considerations are 
less important. In other words, can we use multiple estimates of the 
excess FR---from roughly equivalent spatial locations and for similar 
cluster types---to estimate the mean-squared rotation angle, as defined 
by Crusius-W\"atzel et al.\,  (1990) and by Goldshmidt \& Rephaeli (1993). 
In particular, can we quantify the uncertainties that emerge solely from 
statistical considerations in such estimates of field strengths?
 This quantification has never before been done in this context.

    We begin by defining an ``estimator'' of the mean-squared excess
Faraday rotation angle $\cal E$, namely the arithmetic average
\begin{equation}
{\cal E}= {1 \over N} \sum_{i=1}^N
\chi_i^2\quad.
\end{equation}
Since we have assumed a statistically homogeneous sample, we will assume
for $i=1, ..., N$ that
\begin{eqnarray}
\left\langle \chi_i\right\rangle
&=& 0\cr \left\langle \chi_i^2 \right\rangle
&=&
\sigma^2
\end{eqnarray}
where $\sigma^2$ defines the variance of the process. Further, we invoke
the Central Limit Theorem and the Feller-Lindberg Theorem (as noted above)
so that $\chi$ can be assumed to be Gaussian distributed or
${\cal N}\left( 0, \sigma^2 \right)$. Accordingly, we find that
\begin{equation}
    \left\langle {\cal E} \right\rangle =
{1\over N} \sum_{i=1}^N \left\langle
\chi_i^2 \right\rangle = \sigma^2\quad.
\end{equation}

    As an {\it error estimate}\/ for this process, we define the mean 
squared variation in the estimator $\cal E$ according to
\begin{equation}
    \left\langle \left( {\cal E} - \sigma^2
\right)^2 \right\rangle = \left\langle {\cal
E}^2 \right\rangle - \sigma^4  =
\left\langle {1 \over N^2} \sum_{i,j=1}^N
\chi_i^2 \chi_j^2 \right\rangle - \sigma^4 \quad.
\end{equation}
Taking this further, we find that
\begin{eqnarray}
\left\langle \left( {\cal E} -
\sigma^2
\right)^2 \right\rangle &=& {1\over N^2}
\sum_{i=1}^N \left\langle \chi_i^4 \right\rangle+ {1\over N^2}
\sum_{i\ne j}^N \left\langle \chi_i^2 \chi_j^2
\right\rangle - \sigma^4 \cr &=& {3 \sigma^4
\over N} + {N-1 \over N} \sigma^4
-\sigma^4 \cr &=& {2\over
N}\sigma^4\quad.
\end{eqnarray}
Therefore, we conclude that the RMS uncertainty in this estimator of the
excess FR
\begin{equation}
    \left\langle \left( {\cal E} - \sigma^2
\right)^2 \right\rangle^{1/2} =
\sqrt{2\over N} \sigma^2 = \sqrt{2 \over N}
\left\langle \chi^2 \right\rangle \quad.
\end{equation}
We obtain what may appear as a surprising result, that the uncertainty in
our estimate of the mean-squared excess FR, our estimator $\cal E$, will
generally be comparable to and not much smaller than the estimate itself!
For example, in the data presented by CKB, one cannot find more than
2 data points sharing the same impact parameter---for $N=2$, the
uncertainty above is the same as the estimate itself.
If all 16 clusters provided data with commensurate impact 
parameters -- which unfortunately is not the case -- then the uncertainty 
in the estimate would remain at 35\% of the estimate. 
While CKB provide 27 data, only 12 correspond to the background sources 
that are the basis of our present model.
  The value of $N$ that we employ -- whether we choose 12 or 16 or 27 is 
largely immaterial; the uncertainty in the estimate would remain between 27\% and 
41\% of the estimate itself.

Currently, a number of researchers have taken important steps in performing
the observations to manage instrumental sources of error as well as other
astronomical---yet contaminating---sources of FR. Nevertheless, purely
statistical effects such as those described above provide an absolute
barrier to the accuracy that can be obtained. In order to clarify further
this dilemma, it is instructive to determine the probabilistic distribution
function for $\cal E$ as a function of $N$, beginning with the special case
$N=2$ that arises naturally from the survey of CKB.

    We assume that the distribution for any observed excess FR $\chi_i$,
$i=1, ..., N$, is normal, with variance $\sigma^2$. This can be expressed
mathematically through $\chi_i$'s probability distribution function
$p\left( \chi_i \right)$, namely
\begin{equation}
    p \left( \chi_i \right) = {1 \over \sqrt{ 2
\pi \sigma^2 }} \exp \left\lbrack - {\chi_i^2
\over 2 \sigma^2 } \right\rbrack \quad.
\end{equation}
Consider, first, the special case $N=2$ which has the joint probability
distribution function
\begin{equation}
p \left( \chi_1 \right) p \left( \chi_2
\right) = {1 \over \sqrt{ 2
\pi \sigma^2 }} \exp \left\lbrack -
{\chi_1^2
\over 2 \sigma^2 } \right\rbrack \times {1
\over
\sqrt{ 2
\pi \sigma^2 }} \exp \left\lbrack - {\chi_2^2 \over 2 \sigma^2 }
\right\rbrack\quad;
\end{equation}
meanwhile, we wish to determine the probability distribution function for
obtaining
\begin{equation}
{\cal E} ={\chi_1^2 + \chi_2^2 \over
2}\quad,
\end{equation}
namely $f_2 \left( {\cal E} \right)$, where the subscript 2 denotes the
particular value of $N$ in use. To do this, we use the method of
generating functions (Feller 1968). Consider the Fourier transform
$\tilde f_2 \left( k \right)$ of this distribution function, namely
\begin{equation}
\tilde f_2 \left( k \right) \equiv
\int_{-\infty}^\infty \exp \left\lbrack
ik {\cal E} \right\rbrack \,f_2
\left( {\cal E}
\right) \, {\rm d}{\cal E}
\end{equation}
Given the probabilistic meaning of $f_2 \left( {\cal E }\right)$, it
follows that we can express $\tilde f_2$ according to
\begin{equation}
    \tilde f_2 \left( k \right) =
\left\langle \exp
\left\lbrack ik {\cal E}
\right\rbrack \right\rangle\quad.
\end{equation}
This in turn we can therefore equate with $\tilde f_2 \left( k \right)$
obtained from $\chi_1$ and $\chi_2$, namely
\begin{eqnarray}
\tilde f_2 \left( k \right)
&=&\left\langle
\exp
\left\lbrack ik {\chi_1^2 + \chi_2^2 \over
2}
\right\rbrack \right\rangle\cr& =&
\int_{-\infty}^\infty
\int_{-\infty}^\infty \exp
\left\lbrack ik {\chi_1^2 + \chi_2^2 \over
2}
\right\rbrack\, {1 \over 2\pi \sigma^2}
\exp \left\lbrack -
{\left(
\chi_1^2+\chi_2^2\right)
\over 2 \sigma^2}\right\rbrack \, {\rm d}
\chi_1\,{\rm d}\chi_2\cr
&=& {1 \over 1 - ik \sigma^2}\quad,
\end{eqnarray}
where the last result was obtained by the usual conversion of the double
integral to polar coordinates.

    The calculation is completed by inverting the Fourier transform, namely
\begin{equation}
    f_2 \left( {\cal E} \right) ={1 \over
2\pi} \int_{-\infty}^\infty
\exp\left\lbrack
-ik{\cal E}\right\rbrack\,\tilde f_2
\left( k \right) \,{\rm d}k= {1\over
\sigma^2}
\,
\exp \left\lbrack - {{\cal E
}\over\sigma^2}\right\rbrack\quad,\quad{\rm
for} \quad {\cal E} \ge 0
\quad,
\end{equation}
which was obtained by completing the $k$-contour in the lower-half
$k$-plane around the simple pole at $-i/\sigma^2$. We observe that the
integral $\int_0^\infty f_2 \left( {\cal E} \right)\,{\cal E} = 1$,
verifying that $f_2\left( {\cal E} \right)$ satisfies the properties of a
probability distribution function. To further emphasize this point, we
show in Fig.\ 1 this probability distribution function for the
mean-squared FR excess (gray line) on a frequency histogram showing the
outcome of a Monte Carlo experiment with a sample size of 10,000. (The
variance $\sigma^2$ was assumed to be 1, and $f_2 \left( {\cal E}
\right)$ was suitably scaled to conform with the histogram binning. The
mean value is located at ${\cal E} = 1$.)

\begin{figure}
\plotone{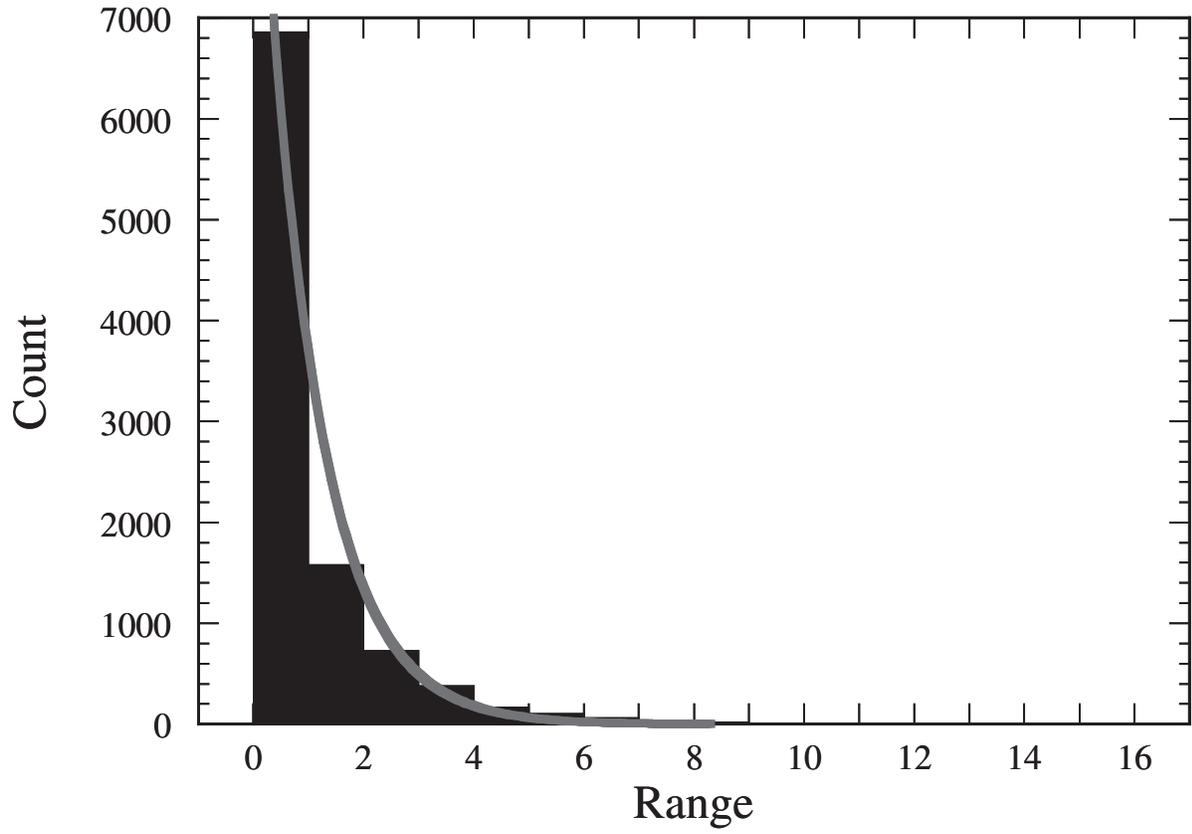}
\figcaption{Theoretical distribution (gray line) superposed on frequency
histogram for 10 000 event simulation of observations with $N=2$ Faraday
excess rotation measurements.}
\label{fig1}
\end{figure}

The purpose of this demonstration is to show that the uncertainty in the
mean-squared Faraday excess, for the best case obtained in the CKB
survey, remains comparable with the result itself. It is manifestly
clear that the uncertainty in this plot, i.e., its width, is
commensurate with its mean value.

As a final illustration of the large measure of uncertainty, and
motivated by the use by  CKB of 16 Abell clusters, we consider the case
$N =16$. The situation is similar to that encountered earlier, except that
\begin{equation}
{\cal E} ={\chi_1^2 + ... + \chi_{16}^2 \over 16}\quad,
\end{equation}
and that the joint probability distribution is the product of 16
Gaussians
\begin{equation}
p_{16} \left( \chi_1, ..., \chi_{16} \right) =  \left\lbrace { 1 \over
\sqrt{2 \pi \sigma^2}}
\right\rbrace^{16}
\exp
\left\lbrack - {\chi_1^2 + ... + \chi_{16}^2 \over 2 \sigma^2}
\right\rbrack\quad.
\end{equation}
Similarly, we observe that
\begin{eqnarray}
\tilde f_{16} \left( k \right)
&=&\left\langle
\exp
\left\lbrack ik {\chi_1^2 + ... +
\chi_{16}^2
\over 16}
\right\rbrack \right\rangle\cr& =&
\int_{-\infty}^\infty \cdots
\int_{-\infty}^\infty \exp
\left\lbrack ik {\chi_1^2 + ... +
\chi_{16}^2
\over 16}
\right\rbrack\, \left\lbrack{1 \over 2\pi
\sigma^2}\right\rbrack^8\,
\exp \left\lbrack -
{
\chi_1^2+ ... +\chi_{16}^2
\over 2 \sigma^2}\right\rbrack \, {\rm d}
\chi_1\cdots{\rm d}\chi_{16}\quad.\cr
\end{eqnarray}
The latter integral can be factored into 8 double integrals, each having
the form seen in Eq.\ (14), thereby giving
\begin{eqnarray}
\tilde f_{16} \left( k \right)
&=&
\left\lbrace \int_{-\infty}^\infty
\int_{-\infty}^\infty \exp
\left\lbrack ik {\chi_1^2 + \chi_2^2 \over
16}
\right\rbrack\, {1 \over 2\pi \sigma^2}
\exp \left\lbrack -
{\chi_1^2+\chi_2^2
\over 2 \sigma^2}\right\rbrack
\, {\rm d}
\chi_1\,{\rm d}\chi_2\right\rbrace^8\cr
&&\cr
&=&{1 \over \left( 1 - i{k
\sigma^2\over 8} \right)^8}\quad,
\end{eqnarray}
where the role of $k$ has been replaced by $k/16$ and from which we obtain
\begin{eqnarray}
    f_{16} \left( {\cal E} \right) &=&{1 \over
2\pi} \int_{-\infty}^\infty
\exp\left\lbrack
-ik{\cal E}\right\rbrack\,\tilde f_{16}
\left( k \right) \,{\rm d}k=
{1 \over
2\pi} \int_{-\infty}^\infty
{\exp\left\lbrack
-ik{\cal E}\right\rbrack
\over \left(1 - i{k \sigma^2 \over 8} \right)^8}
\,{\rm d}k \quad{\rm
for} \quad {\cal E} \ge 0\cr&&
\cr&=&
\left( {8 \over \sigma^2}\right)^8 {1\over 2 \pi} \int_{-\infty}^\infty
{\exp \left( - i k {\cal E} \right) \over \left( k +
i{8\over\sigma^8} \right)^2}
\, {\rm d} k ={8
\over 7!\,
\sigma^2}
\,\left( {8\,{\cal E}\over
\sigma^2}\right)^7
\exp \left\lbrack - {8{\cal E
}\over\sigma^2}\right\rbrack\quad.
\end{eqnarray}
The latter integral was evaluated by performing a contour integration in
the lower-half plane and was evaluated using the calculus of residues at
the compound  pole at $-i8/\sigma^2$. It is easy to check that
$\int_0^\infty f_{16} \left( {\cal E} \right)\,{\cal E} = 1$. We
show in Fig.\ 2 this probability distribution function for the mean-squared
FR excess (gray line) on a frequency histogram, once again superposing
the result on a 10,000 sample Monte Carlo simulation. As pointed out
before, the uncertainty in the mean-squared Faraday excess $\approx
35\%$ of the mean-squared estimate.

\begin{figure}
\plotone{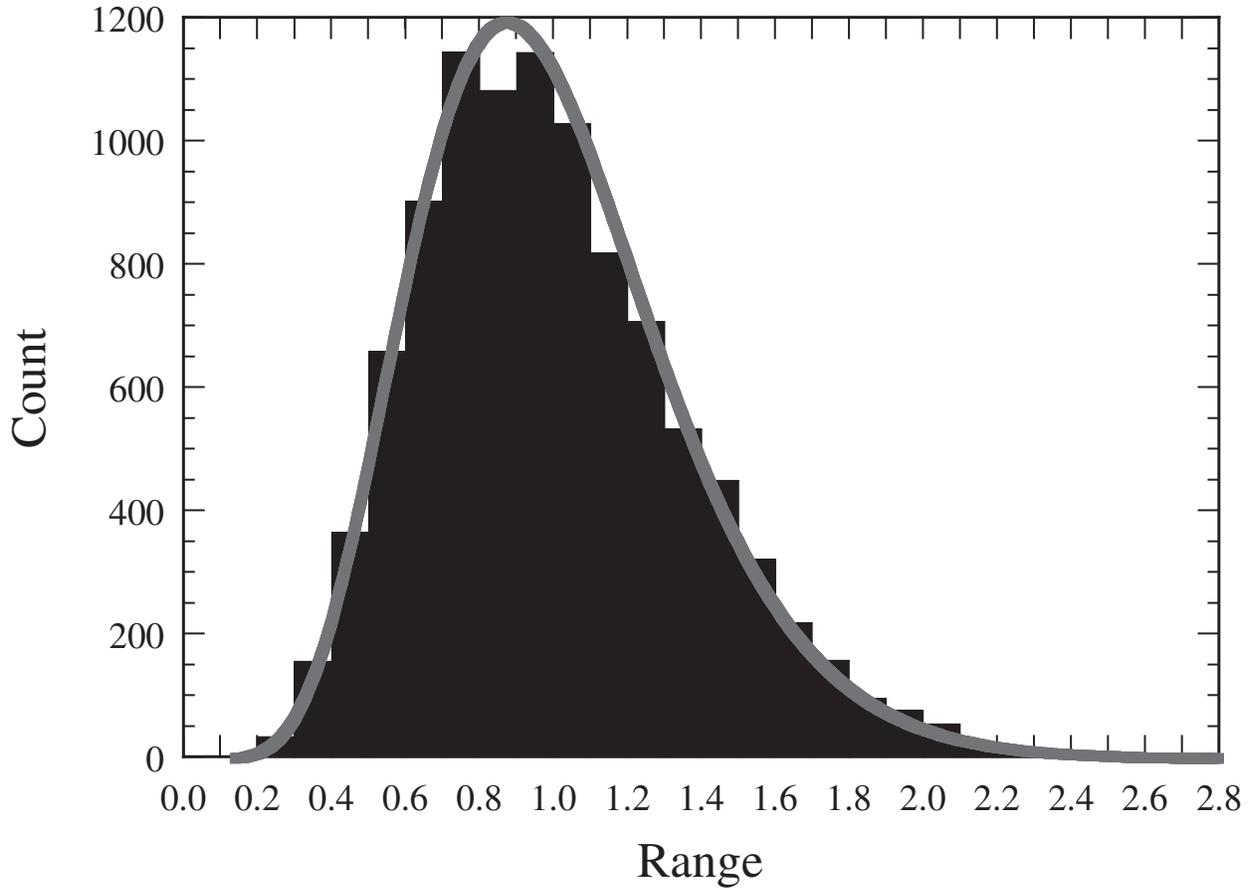}
\caption{Theoretical distribution (gray line) superposed on frequency
histogram for 10 000 event simulation of observations with $N=16$ Faraday
excess rotation measurements.
\label{fig2}}
\end{figure}

We conclude that statistical limitations alone---without
reference to any physical attributes involved in the measurement  of
FR---render the determination of magnetic fields by rotation
measure-based techniques highly uncertain, with the uncertainty in the
estimate being essentially the same size as the estimate itself. In
effect, this result is the outcome of the ``statistics of small
numbers'' (Newman et al.\, 1989), a not uncommon problem in astrophysics.
Now, we wish to address some of the probabilistic or
statistical issues that emerge from spatial heterogeneity in the fields.

\section{Stochastic Field Coherence Effects}

  We explore here the role of spatial inhomogeneity in producing
uncertainty in the estimation of IC magnetic fields. Earlier
theoretical work by  Crusius-W\"atzel et al.\, (1990) and Goldshmidt
\& Rephaeli (1993) considered the role of magnetic turbule size which
they equated with the (reversal) coherence length, $\ell$, using
straightforward random walk arguments. However, we will show, even if the
mean value $\left\langle\ell\right\rangle$ is well-constrained in a given
cluster, that its variability between clusters produces
systematic effects that strongly influence our ability to estimate
the magnetic field strength in the IC medium. The
variation around the mean of some random quantities can
systematically---i.e., in one direction---alter inferences of observed
features, a  point made by Newman et al.\, (1995) in the context of radiative
transfer.  A related phenomenon is at work here, which we presently show.

In the paper by CKB, field strengths were inferred from
their 16 Abell cluster survey assuming either a uniform slab geometry or
random magnetic fields with fixed turbule size $\ell = 10$ kpc, focusing
especially on the latter case. To obtain further insight into this
problem, we performed a Monte Carlo simulation corresponding to their
observations. We assumed the magnetic field amplitudes---but not
directions which were assumed to be random and isotropic---and that
the thermal electron density ($n$) have a (commonly assumed) King
profile,
\begin{equation}
n_t \left( r \right) = {n_0 \over \left(1 + {r^2 \over r_c^2}
\right)^{{3 \over 2} \beta}}\;, \; \quad r < R_d \quad.
\end{equation}
Here, $r_c$ is the gas core radius whose value is in the range
$1/4 - 1/3$ Mpc in nearby clusters; $\beta$ is an empirical
coefficient with typical values between 1/2 and 3/4, and
$R_d$ represents the outer limit to this density distribution,
usually taken to be $\sim 10\ r_c$.

The unknown large scale profile of the locally averaged magnetic field
is likely to be related to the gas density, particularly so if---as
is likely---the fields were ejected from cluster galaxies together
with metal-enriched gas. In this case the radial variation of the field
strength can be assumed to vary as $n \left( r \right)^q$, with $q$
between 1/2 and 2/3, depending on whether magnetic energy or flux is
conserved in this process (Rephaeli 1988). We can then write for
the large scale variation of the field projected on the line-of-sight
\begin{equation}
B_\parallel \left( r \right) = {B_0 \over \left(1 + {r^2 \over r_c^2}
\right)^{{3
\over 2} \beta q}} \;, \; \quad r < R_d \quad.
\end{equation}
where $B_0$ is the central value of the field.
Also of interest is the limiting case of (essentially) identical field
and gas profiles, $q=1$.
Accordingly, we consider values of $q$ between 1/2 and 1. In our numerical
estimates we use the central  values $3 \times 10^{-3}$ cm$^{-3}$, and 1
$\mu$G for the electron density and magnetic field, respectively.
Finally, we assume that a characteristic value of $\ell$---the
coherence length or magnetic turbule size---in a given cluster varies in
the range $1 - 40$ kpc among the set of clusters, with a mean value of
10 kpc, corresponding  to the selection made by CKB.

In order to suitably emulate the observations of CKB, we assumed that
we were simulating observations of 16 Abell clusters with data derived
from these, as shown in their tabulation of observational parameters.
Just as the 27 observed data points were associated with their respective
Abell clusters, we assumed that each of our simulated datum were associated
with their respective and equivalently grouped hypothetical clusters.
Further, we employed the ``impact parameters'' listed in Table 1 of CKB
so that our numerical experiment could qualitatively behave in the same
way as the observations. Our hypothetical clusters were parametrized using
$r_c$, $R_d$, $\beta$, and $q$, as well as $\left\langle\ell\right\rangle$,
which we fixed for each cluster. For each impact parameter $b$, we 
estimated
$\chi$  by a numerical integration of Eq.\ (1), more precisely by using 
CKB's
formula (1), namely
\begin{equation}
RM = 811.9 \, \int_0^L n_t\,B_\parallel\,{\rm d} \ell \quad {\rm rad\
m^{-2}}\quad.
\end{equation}
As a simple device for introducing the randomness associated with magnetic
field orientations, we randomly selected the {\it sign}\/ of the magnetic
field computed from Eq.\ (22) along each line-of-sight segment of length
$\ell$.
Our simulation results are presented in Fig.\ 3 where RM values are
shown against impact parameter for four hypotethical samples (each with
16 clusters). The distributions of physical parameters among these
clusters are as follows: We assume
that $r_c$ is uniformly distributed between 250 and 500 kpc,
$R_d \equiv 10 \, r_c$, $\beta$ is uniformly distributed between 1/2
and 3/4, and $q$ is uniformly distributed between 1/2 and 1.
The four samples differ only in the assumed range of values of
$\ell$. In order to allow as close a comparison as possible with the
observational results of CKB, in our first sample we assumed
$\ell = 10$ kpc in all the clusters. Results of this simulation,
shown in panel (a), closely mimic those of CKB, and this panel is
qualitatively similar to their Figure (1) (open circles between 0
and 1500 kpc). Because of the very similar range of RM values in this
panel and their figure, we use this panel as a `baseline' case in the
comparison of results shown in the other three panels.

\begin{figure}
\plotone{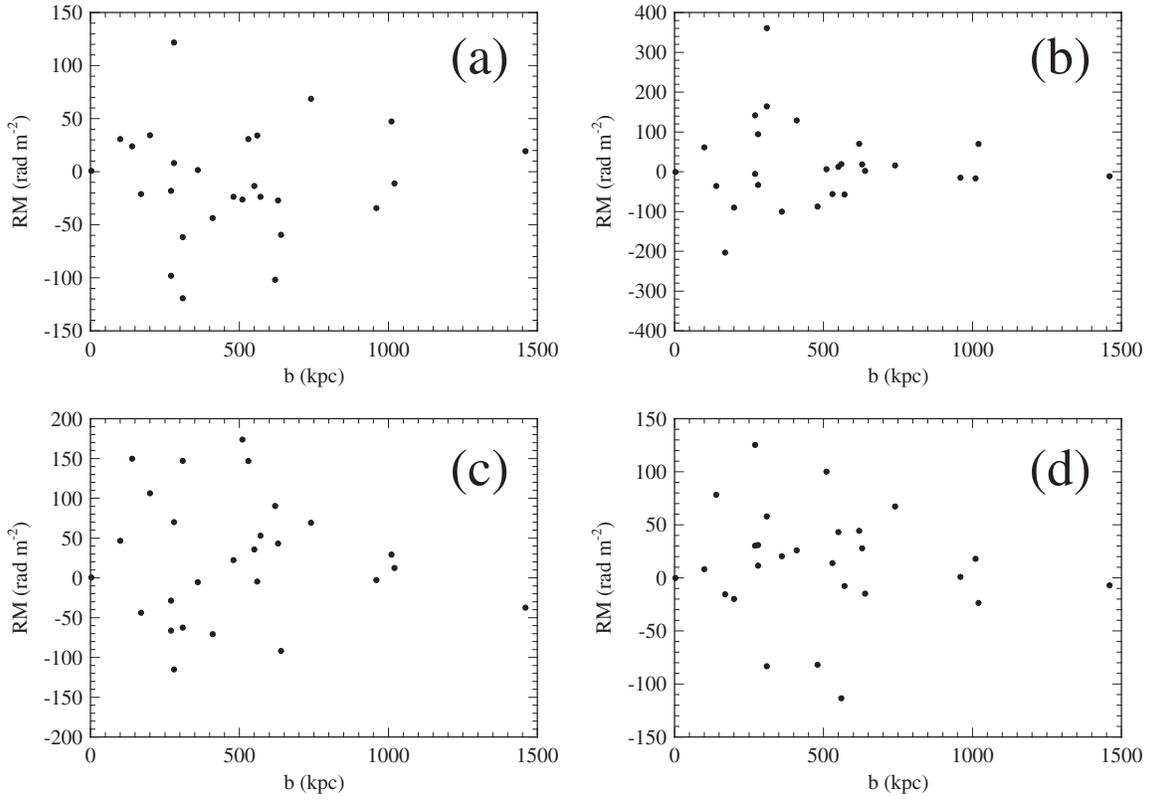}
\caption{Rotation measure plotted as a function of source impact
parameter for our four simulations. See text for details.
\label{fig3}}
\end{figure}

Before we do so we note that an absolute comparison between
our results and those of CKB is somewhat uncertain since we
use a realistic density profile and a range of values for
the gas core radius. Nonetheless, we do not expect that these
differences alone can account for the factor of $\sim 5$ larger
field deduced by CKB from their observed range of RM values, the
same range we have simulated with a central value of just 1 $\mu$G.

Taking $\ell$ to be uniformly distributed between 1 and 40 kpc
yields the range of RM values shown in panel (b). Comparing panel
(b) with panel (a), we see that the RM axis has expanded by
nearly a factor of 3. What has happened here is that the simulated
clusters having relatively large values of coherence length, i.e.,
$\ell > 10$ kpc, contribute larger excess FR angles and this produces
the greater scatter. Had this been an actual set of observations---and
had we taken $\ell= 10$ kpc---we would have inferred a
value of $B_0$ that was a factor of 3 smaller than the fiducial field
assumed in the simulation (1 $\mu$G). The contrast between these
two panels makes it clear how sensitive inference can be to the
{\it assumed}\/ value of $\ell$.

    We use two other distribution functions for $\ell$ in panels (c)
and (d). In (c) we assume that $\ell$ is normally distributed
with a mean of 20.5 kpc and a standard deviation of of 9.75 kpc, so that
2 standard deviations would span the range from 1 to 40 kpc. In (d) we
use a log-normal distribution centered at $6.32$ kpc (the square root of
1 kpc $\times$ 40 kpc) with a variance (in the logarithm) of 2.51
(corresponding to the previous normally distributed case). The results
shown in panels (c) and (d) again reveal the sensitivity to the
distribution in $\ell$. In the normally distributed case, the generally
higher values of $\ell$---compared with the ``baseline'' represented in
panel (a)---result in generally larger values of RM. Again, if these were
real observations, we would be obliged to infer a lower magnetic field,
by nearly a factor of 2. However, in panel (d), the use of a log-normal
distribution puts a heavy weight on generally lower values of
$\ell$---the distribution was centered at 6.32 kpc. Had these been
observations, we would have inferred a larger value for the
magnetic field. At this stage in our investigation of the IC medium, it
is  not clear which (if any) of these panels is more representative of
prevailing conditions. Accordingly, the likely physical variability
of $\ell$ from one cluster to another makes an uncertainty in inferred
fields of a factor of 2 highly probable.

    As a final note in our discussion of the role of variability of 
coherence
length and turbule sizes, we need also mention the likely variability of
$\ell$ within a given cluster. Indeed, we are motivated by the
expectation that the variability of the coherence length $\ell$ around
its mean value $\left\langle\ell\right\rangle$ within a given cluster
will be greater than the variability of the mean coherence length between
clusters. Using simple random walk arguments, it is easy to show that the
mean-squared Faraday excess satisfies
\begin{equation}
\left\langle \chi^2 \right\rangle \propto \left\langle \ell^2 \right\rangle
\end{equation}
where the right hand side of the latter equation refers to the
distribution of cell sizes. However, the Schwarz inequality applies in
this situation; consider, therefore,
\begin{equation}
\left\langle \left( \ell - \left\langle\ell\right\rangle
\right)^2\right\rangle =
\left\langle \ell^2 \right\rangle - {\left\langle\ell\right\rangle}^2
\ge 0 \quad ,
\end{equation}
with the equality satisfied only if $\ell$ is a constant, with no random
variation.
    Thus, it follows that the use of a non-constant $\ell$ in our
simulations of each of the clusters would have produced a systematically
larger RM, thereby implying a smaller field when normalized to a given
measured value of RM.

    As a simple illustration of the significance of this result, suppose
that the turbule sizes are Poisson distributed with mean $\ell_0$ in the
IC medium, that is
\begin{equation}
p \left( \ell \right) = {1\over\ell_0} \exp \left( -{\ell\over\ell_0
}\right)\quad,
\end{equation}
from which we obtain that
\begin{equation}
\left\langle \ell^2 \right\rangle = 2 \ell_0^2 {\rm \quad and \quad}
\left\langle \ell \right\rangle=\ell_0 \quad.
\end{equation}
    We note that a Poisson distribution of turbule sizes would introduce a
factor of 2 into Eq.\ (24), that is, the right hand side should be a
factor of 2 larger than would be obtained by assuming turbules of uniform
size and that this would necessarily change the magnetic field estimate.

    While the preceding illustration provides a hint as to the nature of the
problem, deeper insight can be obtained by considering the possible dynamo
origin of magnetic fields in galaxy clusters (Ruzmaikin et al.\, 1989, Zeldovich
et al.\, 1988). Suppose that fully-developed, isotropic
turbulence applies and that the distribution of turbule sizes can be
related to a Kolmogoroff turbulent spectrum (Ferraro and Plumpton 1966).
This could be the outcome of the injection of galactic fields into the IC
medium, thereby ``stirring'' magnetized fluid contained within, resulting
in the turbulent transport of energy from these largest scales to the
smallest, dissipative scales. We will not consider in the present
discussion the possibility that such turbulent activity could be further
complicated by the emergence of rope-like structures, such as those
mentioned by Ruzmaikin et al.\, (1989) and Zeldovich et al.\, (1988).
    The Kolmogoroff spectrum projected along a line-of-sight would have the
characteristic $k^{-5/3}$ form, where $k = 2\pi/\ell$ is the wavenumber
that would characterize a turbule of size $\ell$.
    Accordingly, integrals containing terms like $k^{- 5/3} \, dk$ would
transform into $\ell^{- 1/3} \, d\ell$ by virtue of $\ell = 2\pi/k$.
    We now assume that  the Kolmogoroff spectrum has lower and upper
length-scale cut-offs, $\ell_1$ and $\ell_2$, respectively---these
emerge from presence of dissipation and production ranges for the
turbulence and we shall assume that $\ell_2 \gg \ell_1 > 0$.
    In this environment, average values of any quantity $f \left( \ell
\right)$ can be calculated according to
\begin{equation}
\left\langle f \left( \ell \right) \right\rangle =
{\int_{\ell_1}^{\ell_2} f
\left(\ell \right)\,
\ell^{-1/3} \, {\rm d}\ell \over \int_{\ell_1}^{\ell_2}
\ell^{-1/3} \, {\rm d}\ell} \quad.
\end{equation}
Hence, we find that
\begin{equation}
\left\langle \ell^2 \right\rangle \approx {1\over 4} \ell_2^2 {\rm
\quad and
\quad} \left\langle \ell \right\rangle \approx {2\over 5} \ell_2 \quad,
\end{equation}
and
\begin{equation}
\left\langle \ell^2 \right\rangle = {25\over 16} \left\langle \ell
\right\rangle^2  \quad.
\end{equation}
    We observe in this physically-relevant derivation that we recover a
factor approaching 2 germane to our estimation of field strengths, and
that the statistic emphasizes the {\it longest} turbule sizes (with a
multiplying factor of 2/5). For the reasonable range of $1 - 40$ kpc,
a representative mean turbule size would be $\sim 2/5 \times 40 = 16$
kpc, rather than the fiducial 10 kpc that has (typically) been used.

    The importance of this discussion regarding the variability of $\ell$
among clusters and inside each cluster is that this variability
introduces further uncertainty, perhaps as much as an additional
factor of 2, in our ability to estimate the prevailing magnetic
fields. This magnetic turbulence-generated issue, together with the
strictly statistical ones discussed in the previous section, introduce
substantial uncertainties in values of the field deduced from FR
measurements.

\section{Radiative Compton-Synchrotron Measures of Magnetic Fields}

Spectral and spatial measurements of the emission due to synchrotron
and Compton energy losses of a population of relativistic \,  electrons can yield
the electron spectral density distribution and magnetic field in the
emitting region, and at least some information on the spatial profiles
of these quantities. Over the large ($\sim 1$ Mpc) regions in clusters
the electron density and magnetic field are expected to vary
significantly, yet because  of lack of adequate spatial information, this
variation was ignored  in all previous estimates of IC magnetic fields
from radio and (NT)  X-ray measurements ({{\it e.g.},\,} RGB, Fusco-Femiano et al.\,
1999). We now want  to assess how such estimates are affected by the
profiles of the
relativistic \, electrons and field.

   Suppose that the electron Lorentz factor ($\gamma$) and radial ($r$)
distributions are (separable and) given by
\begin{equation}
n(\gamma, r) = A \gamma^{-p} g(r),
\end{equation}
and let the profile of the {\it locally-averaged root mean squared}
field (over regions at least the size of $\ell$) be denoted by the
function $b(r)$. The value of the field from radio and X-ray
measurements is then proportional to the ratio (Rephaeli 1979):
\begin{equation}
\zeta = \left\lbrack { \int_0^{R_d} r^2 g(r) dr \over
   \int_0^{R_d} r^2 g(r) b(r)^{p+1 \over 2} dr } \right\rbrack
^{ 2 \over p+1}
\end{equation}
Likely deviation of the value of this ratio from unity introduced
significant systematic uncertainty in the values of $B_{rx}$ that have
previously been derived from radio and X-ray measurements.

To obtain some insight into how these geometrical considerations
influence our estimates of $B_{rx}$, we have computed $\zeta$ for
a variety of field and relativistic \, electron profiles. Of these, perhaps most
appropriate (since we have little information on the field and relativistic \,
electron spatial distributions) is to use the same functional form
as that of the gas (eq. 21), but with a reasonably wide range of field
and electron core radii, $r_B, \, r_e$, respectively. The motivation
for doing so is the expectation that field has likely originated in
the cluster galaxies (Rephaeli 1988). Its spatial distribution
is expected to be closely related to that of the hot gas, namely
wide-spread throughout most of the cluster. The origin of relativistic \,
electrons is less clear; models range from origin in a few
central radio galaxies (Rephaeli 1977), to ongoing shock
acceleration by fast moving spiral galaxies ({{\it e.g.},\,} Bykov et al.\, 2001),
implying a wider spatial distribution. (For more on electron models,
see Sarazin 1999, and Rephaeli 2001.) Now, the observed regions of radio
emission have sizes $\sim 1-2$ Mpc, and with $1/4 - 1/3$ Mpc as a
typical range of gas core radii, it is clear that the field and electron
profiles cannot be very centrally concentrated. On the other hand, it
is also unlikely that either of these profiles---particularly that of
the electrons---is more extended than that of the gas: Hot IC gas is
a major, global component of clusters, whereas magnetic fields and
relativistic \, electrons are less ubiquitously abundant, and require special
conditions to maintain, such as mechanisms for generation or
amplification, and acceleration to balance energy losses,
respectively.

In the table below we list some illustrative values of $\zeta$ computed
for values of $r_B / r_c$ and $r_e / r_c$ in the range of $0.2 - 1$.
While these results are general and do not depend on a specific value
of $r_c$, we consider the lowest relevant values of $r_B$ and $r_e$ to
be $\sim 50$ kpc, values corresponding to the lowest ratios in the
table for  $r_c= 250$ kpc. We have taken the cluster limiting radius
to be $10r_c$, and assumed that the sky-projected radius of the emitting
region is $2r_c$, roughly corresponding to the region sampled by both the
FR and radio measurements. The electron energy spectrum is assumed to be
a power-law with index in the range $p=3-5$, i.e., corresponding to the
range $1-2$ for the radio spectral index, $\alpha$. The best measured
cluster radio spectrum is that of Coma, with $\alpha \simeq 1.34$ (Kim
et al.\, 1991), implying $p \simeq 3.68$.

\begin{table*}[t]
\centerline{The factor $\zeta$ in the expression for $B_{rx}$}
\begin{center}

\bigskip
\bigskip

\begin{tabular}{|cc|cccc|}
\hline
             $r_B/r_c$  & $r_e/r_c$  &       &   $\zeta$  &        &      
\cr

                      &         &   p=3.0  &   p=3.68   & p=4.0  &  p=5.0 
\cr
\hline

             0.2      &  1.0    &  24.0    &   17.9     &  15.8  &  11.4  
\cr
                      &  0.6    &  20.7    &   15.7     &  14.0  &  10.3  
\cr
                      &  0.2    &  13.5    &   10.9     &   9.9  &   7.7  
\cr
             0.6      &  1.0    &   5.2    &    4.7     &   4.5  &   3.9  
\cr
                      &  0.6    &   4.6    &    4.2     &   4.0  &   3.6  
\cr
                      &  0.2    &   3.4    &    3.2     &   3.1  &   2.8  
\cr
             1.0      &  1.0    &   2.9    &    2.7     &   2.7  &   2.5  
\cr
                      &  0.6    &   2.6    &    2.5     &   2.5  &   2.3  
\cr
                      &  0.2    &   2.1    &    2.0     &   2.0  &   1.9  
\cr
\hline
\end{tabular}
\end{center}
\end{table*}

We see that $\zeta$ (which is always $>1$) assumes values
in the interval $\simeq 2 - 24$ for the range of core radii taken here.
Specifically, in the Coma cluster---using the measured radio and (assumed
Compton) X-ray fluxes, and ignoring the unknown spatial profiles
(effectively, taking $\zeta = 1$), RGB deduced a value of $\sim 0.2$
$\rm\mu G$ for $B_{rx}$, which we denote here as $B^{U}_{rx}$. This
fiducial value is to be contrasted with the estimates based on $\zeta
\neq 1$: Had RGB assumed the above field and electron profiles and
parameters, they would have deduced values of $B_{rx}$ in the center of
the cluster which are in the range $\sim 0.4 - 4.8$ $\rm\mu G$.

Having illustrated the impact of spatial profiles of the gas, field and
electrons, we should emphasize that even though these profiles are
realistic, the viable ranges of core radii are likely to be
narrower than considered here. The latter will only be known once
detailed measurements are made of the spatial distributions of the
radio and NT X-ray emission. It is also important to note that the
volume-averaged mean field value is generally lower than the product
$B^{U}_{rx}\times \zeta$ by a factor that depends very much on the
way this average is defined.

\section{Discussion}

We have explored several sources of uncertainty in the estimation of
magnetic field strengths in clusters. The first of these, related to
the measurement of excess FR, is of a purely statistical source and,
in essence, is the outcome of the statistics of small numbers. The
lack of large numbers of clusters that can be adequately probed by
this technique results in an uncertainty in the estimate of the
mean-squared excess FR that is comparable with the estimate itself.
The second of these has a more physical basis, yet is due to the
underlying randomness in the spatial structure of turbulent magnetic
fields. The variability of the
coherence length or magnetic turbule size within each cluster and among
clusters can introduce uncertainties in the field approaching an order
of magnitude. Finally, the traditional use of Compton and synchrotron
emissions to determine IC field strengths requires knowledge of
the unknown field and relativistic \, electron morphologies, introducing yet another
source of uncertainty, one that could easily translate to a factor of
at least a few in the value of the deduced mean field.

A realistic assessment of the above uncertainties is of obvious
relevance to the study of nonthermal phenomena in general, and to
the modeling of the origin of the fields and energetic electron
populations in particular. Mean field values of ${\cal O} (10^{-7})$ G
can be comfortably expected from ejected galactic fields, but
not so if the mean field value is ${\cal O} (10^{-6})$ G. The
implied electron energy density, $\rho_e \propto B^{-(p+1)/2}$,
depends steeply on B. An uncertainty of a factor of a few in B
typically implies more than an order of magnitude difference in
$\rho_e$ (and the related energetic proton energy density). This
too can make all the difference in judging the viability of models
for particle origin and the need for effective re-acceleration
processes in the IC space.

Of course, the range of electron energies implied from the observed
radio spectral range also depends directly on the field value,
though much less steeply ($\gamma \propto B^{-1/2}$), than the
energy density. The lower the field value, the higher is the
energy of the synchrotron emitting electron, and the shorter is its
(dominant radiative) Compton energy loss time ($\propto 1/\gamma$).
A short loss time could mean either that radio and X-ray emission might
be short, essentially transient phenomenon, or that the particles are
continually accelerated in order to sustain these emissions over
cosmological time.

As mentioned in the Introduction, the electron energy spectrum is
likely to extend to energies well below the range probed directly
by radio measurements. Indeed, in the simplest model, if a steady
state is attained then relativistic \, electrons are continually replenished
as they lose energy, first radiatively and then by electronic
(Coulomb) excitations ({{\it e.g.},\,}, Rephaeli 1979). For any reasonable
mean IC magnetic field value, only high energy ($\gamma
 > 10^3$) electrons contribute to the observed radio emission.
Therefore, if the electron energy spectrum is correctly extended to
lower energies (by taking Coulomb losses into account) then this will
not have appreciable impact on our above considerations. If, on the
other hand, an energetic electron population is conjectured in order
to account for the claimed measurement of EUV emission in the Coma
cluster, then extension of this population to higher energies may
in general imply a higher magnetic field ({{\it e.g.},\,}, Sarazin \& Kempner
2000) then our deduced values of $B_{rx}$. However, the proposed
energetic electron models are quite unrealistic due to their very
high energy densities, which are comparable to or even higher than
the gas thermal energy density (Rephaeli 2001, Petrosian 2001).

The CKB core sample consisted of 27 radio sources, only 12 of which
were background sources. The inclusion in the sample of a large
fraction of cluster members introduces an additional uncertainty
which stems from the possibility that these sources have
(statistically) different radio properties from those of the
background sources, and the expectation that the IC pathlength for
a cluster member is on average only half that of a background source.
The issue whether the mixed nature of such a sample affects estimates
of the mean IC field was previously investigated by Goldshmidt \&
Rephaeli (1993) in their re-analysis of the Kim et al.\, (1991) sample.
By repeating the analysis of FR measurements of cluster vs. non-cluster
members, Goldshmidt \& Rephaeli (1993) concluded that the distribution
of rotation angles of cluster members was actually somewhat narrower
than that of non-cluster members, perhaps due to the reduced pathlength
of the former sources. This difference was small enough that estimates of
the field were hardly affected. The issue has to be further explored
before definitive statements can be on the magnitude of the systematic
uncertainty due to a `mixed' radio sample.

The systematic uncertainties we have expanded upon should be
better understood and more explicitly acknowledged whenever values
of the field (and electron energy density) are reported. This is
{\it crucial} when values of $B_{fr}$ and $B_{rx}$ are compared.
In particular, we caution against strong conclusions that are
sometimes drawn on the non-viability of the Compton interpretation
of NT X-ray emission based on the discrepancy between deduced values
of $B_{fr}$ and $B_{rx}$. As we have clearly demonstrated here, the
current status of the measurements, and their associated substantial
uncertainties, render such conclusions premature.

\acknowledgments
We wish to thank Tracy Clarke and Larry Rudnick for helpful discussions. 
W.I.N. was the Belkin Visiting Professor in the Department of Computer
Science and Applied Mathematics of the Weizmann Institute of Science, and
gratefully acknowledges its support. A.L.N. gratefully acknowledges the
support and hospitality of the Department of Astronomy and Astrophysics
at Tel Aviv University where this research was performed. Y.R. acknowledges 
NASA for its support of this project.

\end{document}